# *Presenting particle physics and quantum mechanics to the general public*

J. Strauss (*)

Abstract:

The job of a physicist is to *describe Nature*. General features, hypotheses and theories help to *describe* physical phenomena at a more abstract, fundamental level, and are sometimes tacitly assigned some sort of real existence; doing so appears to be of little harm in most of classical physics. However, missing any tangible connection to everyday experience, one better always bears in mind the *descriptive nature* of any efforts to grasp the quantum. And elementary particles interact in the quantum world, of course. When communicating the world of elementary particles to the general public, the *Bayesian approach* of an ever ongoing updating of the depiction of reality turns out to be virtually indispensable. The human experience of providing a series of increasingly better descriptions generates plenty of personal pleasures, for researchers as well as for amateurs. A suggestive analogy for improving our understanding of the world, even the seemingly paradoxical quantum world, may be found in recent insight into how congenitally blind children and young adults *learn to see*, after having received successful eye surgery.

Presenting particle physics to a general audience is easy and tricky at the same time. The easy part consists of presenting the instrumentation, the gigantic detectors and the hordes of researchers, where every single person performs some necessary tasks, some *snippets* of the whole enterprise. Young researchers get experienced over time, and progressively start to perceive the whole picture, also aided by *'standardized' public talks* that they themselves as well as their supervisors have to deliver, with increasing frequency, since the taxpayers have the right to get informed.

Eventually every *standard talk* reaches a point of trying to illuminate motivations, main results and future outlooks. Here comes the tricky part where the interaction with the public often tends to become less satisfactory. Do Quarks really exist? Are they particles or waves? Does the Higgs field pervade all of space? Are we indeed bathing in Dark Matter? Is a high-energy collision between nuclei a replay of the cosmological Big Bang? (And so forth.) Depending on the very public at least as much as on the skill of the person giving the presentation, the topic of the discussion may shift towards pointing out various mysteries (our evolutionary past as hunters makes us enjoy mysteries), and possibly arriving at some famous *eternal* questions, often with a sigh of regret that the performing scientist is oh so clever, and the poor man or woman in the audience will never be able to understand the exciting conclusions. The feeling that such thoughts seem to be entertained by the public is embarrassing. In this article, a possible way will be suggested to improve the general understanding of particle physics, and the quantum world.

The key to a better intuitive feeling for the physics beyond our everyday experience should not be expected from an ultimate, groundbreaking scientific discovery, but rather from a realistic insight into what science is about: Science is about *description,* and <u>not</u> *explanation* of nature. An example of a scientific question would be '*How does electricity work?*' and not '*What is electricity?*' The researcher is thus freed from any preconceived model about the *true* nature of an observed phenomenon and can put the priority on experimental facts, in spite of possible worries about contradictions: 'Contradiction is not a sign of falsity, nor the lack of contradiction a sign of truth' (Blaise Pascal); here, the quote may stand outside of the original context, but seems to be most useful for our intentions. Immanuel Kant has systematized these issues, and Bertrand Russel has written two entertaining little volumes about epistemology [1]. This article does neither attempt to give an appreciation of the topic, nor to confront B. Russel's work with recent scholarly philosophy.

Russel's '*knowledge of general principles*', which is needed to put empirical findings into context, seems to exist for humans as well as, in some basic form (i.e. without the copious human self-referencing aspects), for any living being and even for many biological macromolecules, and is often denoted as *Emergence* and *Darwinian evolution.*

But all that is just philosophy of science and is quite remote from the daily work in research labs. A researcher enjoys the *pleasure of finding things out* (Richard Feynman), and gets his satisfaction from formulating his findings in a concise form. Inside of the science community, philosophy is superfluous, since everybody knows more or less what is being talked about, and there is virtually no need to be constantly reminded about foundations. It's still useful to be aware of them when contemplating new projects, and also in contacts with people outside the community.

In the course of the last two decades - since entangled quantum states are being prepared and studied in laboratories around the world on a regular basis, and even being commercially used for encryption - questions from the audience are more and more frequently concerned with some famous and popular puzzles of the quantum world [2]. But why is the successful description of macroscopic objects by their masses, positions and velocities supposed to be less puzzling than the description of spatially extended and seemingly disconnected quantum phenomena by a wave function, or by any other tool to be used to calculate quantum probabilities? The question is not meant to alleviate worries about quantum physics, but, more generally, to sharpen our judgment about what *describing* nature means, as opposed to explaining nature.

Before ruminating about the intricacies of elementary particles, it may be useful to consider some insight from another area of research, namely about how our eye-brain system *learns to see* and recognize objects.

How do children learn to grasp our visual world? This is the subject of vigorous research in neuroscience, and one aspect of this question was summarized in a recent article written by Pawan Sinha [3]. His *Project Prakash* is intended to eventually help, by simple and cheap eye-surgery, an estimated 400.000 cataract-stricken, blind children and young adults (some of them well into their 20s) in India. The project has met its first successes.

At the start of *Project Prakash*, there were some nagging doubts whether any visual function could be acquired after treatment for blindness late in childhood: When we open our eyes, the huge collection of pixels of an image is, without any apparent effort, organized into an orderly collection of things. Yet Pawan Sinha and his collaborators have found that the experience of a Prakash child soon after gaining sight is different. The newly sighted exhibits profound impairments. They are unable to organize the many regions of different colors and brightness into larger assemblies. Many features of ordinary objects – the overlapping sections of two squares or adjacent sections of a ball delineated by the lacing on its surface – are perceived as entirely separate objects, not component parts of larger structures. As a particularly striking example, a ball and its shadow are seen as distinct objects.

Interestingly, the confusing soup of pixels is recognized as a meaningful pattern with the introduction of one particular visual cue: motion.

Our brain has learned to recognize an object as the *invariant* of a part of a visual scene under translation and/or rotation. That's how an assembly of 'proto-objects' becomes an object: namely after a successful pattern recognition task of singling out an invariant, which is characterized by many parameters including size and details of its form.

This learning process is an example for the incremental improvement of the *description* (=*understanding*) of the world around us. It may be considered to be an illustration of an ever ongoing *Bayesian updating* [4].

In the Bayesian view, probability measures the personal degree of belief, and the famous Bayes's theorem links the degree of belief in a proposition before and after accounting for new evidence. Imagine a gathering of human agents, all being experts in statistics, confronted with the need to bet on the outcome of an event: The probability of an outcome is not inherent in that event. Different agents, with different beliefs, will in general assign different probabilities. (By the way, the calculus of probability theory is equivalent to the postulate that only *sure loss* is not allowed in an agent's bet [4]. This seems to be quite an intuitive embedding for probability theory, and it is a central paradigm in economic sciences.) The Bayesian interpretation of probability *puts the scientist back into science* [5]. Of course, most of the time the terminus 'scientist' denotes a wider *peer group*.

In the Quantum Bayesian view, the quantum wave function is a construct serving to *describe* quantum probability as judged by a specific observer, rather than an object of reality: Quantum probability depends on the observers *personal* bet, which is of course his or her <u>informed</u> bet ("*hypotheses non fingo*"); quantum theory serves to organize those bets, it is "a theory guiding agents in their interactions with the [quantum] world" [6].

Instead of a Prakesh child who is for the first time confronted with visual cues, imagine a physicist who is for the first time confronted with the quantum realm.

The exemplary physicist will learn to make sense of the quantum world. What are the transformations? What are the invariants?

Having *learned to see*, we are able to recognize various three-dimensional objects as invariants under certain transformations, and those objects may differ in myriads of ways. The phenomena in the quantum realm have no resemblance to that. Quantum phenomena are really simple. There exist just a few invariants. Quantum coherence, meaning that abstract concepts are needed to describe the phenomena, makes them look quite unusual when compared to our everyday experience. But unusual is already their simplicity! We have difficulties to *understand*, which means 'to compare to something we know'. It's amusing to notice, however, that some sophisticated molecules functioning in certain biological settings have already acquired the ability to 'profit' from genuine quantum features. Chlorophyll is a case in point: some of the crucial and spatially separated groups of atoms must be entangled to enable photosynthesis. The apparent *contradiction* of experience with quantum phenomena to the experience with everyday objects *is not a sign of falsity.*

Richard Feynman famously remarked that *nobody understands Quantum Mechanics* [7]. He pointed out the 'analogy and contrast' of elementary particles to bullets and water waves, 'bullets' representing macroscopic objects, and 'water waves' periodic phenomena (by the way, waves are generic, and there seems to be no need for 'proto-waves').

But why struggle with 'analogy and contrast'? Unlike macroscopic objects, quantum phenomena are simple. In elementary particle physics, we only have a few particle varieties. All particles of a given variety are identical.

An energetic, electrically charged and stable hadron passing thru matter leaves an ionization trail, and reveals its energy by depositing it by electromagnetic and hadronic interactions in a dense enough detector material of sufficient depth. This looks like a bullet leaving a trace, and eventually stopping in soil. But unlike a bullet, we are able to acquire only very limited knowledge about the particle. With a "size" of about $10^{-15}$ m, it produces on average 20 observable primary ionization pairs per $10^{-3}$ m track length in gas, and eventually deposits its total energy in about 1 m of iron. Heisenberg uncertainties are too small to be noticed in such an experimental set-up, and there is no need to describe the free particle in the asymptotic region (i.e. far away from where it was produced) by a wave function.

By scaling up the hadron to the size of a real bullet of a diameter of $10^{-2}$ m, about one signal every $10^9$ m could be detected along its path, which is three times the distance to the moon. Hence it is *not appropriate* to think of an elementary particle as a bullet. The particle is a much simpler entity. In terms of its time- and length-scales, it left the interaction region where it was produced a very long time ago, and a very long spatial distance away. But by asking the right question about its momentum, energy and quantum numbers, it can reveal some information about the interaction, where it was produced in company with other particles.

Interacting elementary particles have to be described by the full-fledged formalism of operator-fields acting on infinite dimensional Hilbert space. We clearly use short cut phrasing when we talk about interacting 'particles' and 'fields'.

The various 'fields' (or 'particles'), and the scheme of how the fundamental fermions and gauge bosons are organized in the *Standard Model*, and how they interact, is the result of an astonishing, decades-long progress of adapting quantum mechanics to the world of quarks, leptons, gauge bosons, and the Higgs boson. (In the following, the Strong sector with quarks and gluons will not further be commented on.)

The success story began some time ago with the formulation of the theory of Quantum electrodynamics (QED), the quantum version of electromagnetism. QED implies, that the vacuum - when investigated at small space-time scales - is in fact full of fluctuations, which can even be observed at macroscopic scales (*Casimir effect*) by measuring the force between two uncharged, very close parallel conducting plates (in reality a very tricky measurement!). The description of the quantum version of electromagnetism with operator-fields indeed turns out to be very successful. One might almost be tempted to reify those fields.

Imagine for a moment, that a Casimir force could not be detected, despite of sufficient experimental sensitivity. Somebody doubting the *reality* of operator-fields might be enchanted, but falsely so, because the absence of the effect would be a statement about a contradiction of nature with its description by QED. An improved, revised description would have to be invented.

After early attempts by Enrico Fermi, the description of the Weak interactions experienced various ups and downs. The Intermediate vector bosons $W^+$, $W^-$ and $Z^0$ (witnessing the piecing together of the weak and the electromagnetic as the electro-weak interactions) came at the right time to regularize the electro-weak sector, and the discovery of a Higgs boson provides for a version of the breaking of the electro-weak symmetry to result in nonzero masses of the W and Z bosons. Now it remains to be seen how the mass of the Higgs boson avoids infinite quantum corrections, and how it is stabilized at its experimentally measured value. The answer will be a further elaboration of the quantum fields paradigm; or its breakdown, signaling the need for some completely new way of understanding (=describing) elementary particles. Note that quantum gravity still presents a major challenge.

Finally, I wish to come back to the initial question of 'How to present particle physics to the general public?' Some hints at the basics of the *scientific method* are usually fruitful. I try to establish a vivid contact with the audience and to encourage the visitor to voice his or her own perception. Various people have various views and expectations. The description of nature by humans is a human enterprise. Dogmas don't exist in science, but consensus has to be established.

A few random and typical questions have been posed in the introductory paragraphs, and in the conclusion some simple answers may be in place: Quarks 'really' exist, if we talk in terms of the Standard Model of fundamental fields; they are neither particles nor waves, but genuine quantum phenomena. We may talk about the Higgs field, if we have in mind an operator field, which helps to bestow masses to the W/Z bosons without destroying the nice properties of their interactions. About dark matter, we don't know enough yet. A high-energy collision between nuclei is certainly not a replay of the cosmological big bang. (And so forth.) It seems to me that pop science gibberish often rules in those spheres, which are quite remote from direct experience of most of us, and which will never have any *practical* relevance to our daily lives.

It's definitely easier to talk about particle detectors and to understand their functioning, eventually to ponder questions of scientific, technological and economical feasibilities of small and big gadgets.

Science entertainment, though not every scientist's cake, is very much appreciated by the public. I once came across a group of people who liked to listen to physicists, even if they did not understand a single word; it was just for the performance, and for being exposed to so many termini not used in everyday conversations. May be that's similar to reading *Finnegans Wake*? To *teach curiosity*, which is at the root of any scientific endeavor, and to communicate *The Pleasure Of Finding Things Out* is worth any effort, and is met with high esteem by the susceptible.

Suggestions and encouragements by Michael Dittmar, Peter Feistel, Chris Fuchs, Samo Kupper, Fritz Vogt and Estella Weiss-Krejci are gratefully acknowledged.

(*) The author is a retired experimental physicist. He was a member of the Institute of High Energy Physics in Vienna, Austria, and he spent over ten years at the European Research Lab CERN in Geneva. The remarks collected in this opinion piece arose from watching, and occasionally participating in reach-out activities of the high-energy physics community.

emails: Josef.Strauss@assoc.oeaw.ac.at, Jozko.Strauss@aon.at